\documentclass[10pt, a4paper, twocolumn, twoside]{article} 

\usepackage[english]{babel} 

\usepackage{microtype} 

\usepackage{amsmath,amsfonts,amsthm} 

\usepackage[svgnames]{xcolor} 

\usepackage[hang, small, labelfont=bf, up, textfont=it]{caption} 

\usepackage{booktabs} 

\usepackage{lastpage} 

\usepackage{graphicx} 

\usepackage{enumitem} 
\setlist{noitemsep} 

\usepackage{sectsty} 
\allsectionsfont{\usefont{OT1}{phv}{b}{n}} 

\usepackage{geometry} 

\usepackage{float}

\usepackage{wrapfig}

\usepackage[T1]{fontenc} 
\usepackage[utf8]{inputenc} 

\usepackage{XCharter} 

\usepackage{pdfpages}
\usepackage{subfigure}

\usepackage{fancyhdr} 
\fancypagestyle{firstpage}{ 
	\fancyhf{}
	\fancyfoot[LE,RO]{\footnotesize Page \thepage\ of \pageref{LastPage}}
}

\definecolor{CCblue}{rgb}{0.176, 0.239, 0.34 }
\definecolor{CCyellow}{rgb}{0.98, 0.79, 0.406}

\definecolor{ultramarine}{RGB}{0,32,96}
\newcommand{\authorstyle}[1]{{\large\usefont{OT1}{phv}{b}{n}\color{CCblue}#1}} 

\newcommand{\institution}[1]{{\footnotesize\usefont{OT1}{phv}{m}{sl}\color{Black}#1}} 

\usepackage{titling} 

\newcommand{\HorRule}{\color{CCblue}\rule{\linewidth}{1pt}} 

\pretitle{
	\vspace{-65pt} 
	\HorRule\vspace{10pt} 
	\fontsize{32}{36}\usefont{OT1}{phv}{b}{n}\selectfont 
	\color{ultramarine} 
}

\posttitle{\par\vskip 15pt} 

\preauthor{} 

\postauthor{ 
	\vspace{10pt} 
	\par\HorRule 
	\vspace{-15pt} 
}

\usepackage{lettrine} 
\usepackage{fix-cm}	

\newcommand{\initial}[1]{ 
	\lettrine[lines=3,findent=4pt,nindent=0pt]{
		\color{CCyellow}
		{#1}
	}{}%
}

\usepackage{xstring} 

\newcommand{\lettrineabstract}[1]{
	\StrLeft{#1}{1}[\firstletter] 
	\initial{\firstletter}\textbf{\StrGobbleLeft{#1}{1}} 
}

\usepackage[backend=bibtex,style=authoryear,natbib=true]{biblatex} 

\usepackage[autostyle=true]{csquotes} 

\title{A Technological Perspective on Misuse of Available AI} 

\author{
	\authorstyle{
		Lukas Pöhler\textsuperscript{1,2}, 
		Valentin Schrader\textsuperscript{1,2}, 
		Alexander Ladwein\textsuperscript{1,2} 
		and Florian von Keller\textsuperscript{1,3}
	} 
	\newline\newline 
	\textsuperscript{1}\institution{
		ConsciousCoders (https://www.consciouscoders.io), Munich, Germany}\\ 
	\textsuperscript{2}\institution{
		Technical University of Munich, Munich, Germany}\\ 
	\textsuperscript{3}\institution{
		Ludwig-Maximilians-Universität München, Munich, Germany}\\ 
}

\date{} 

\begin{document}


\maketitle 


\thispagestyle{firstpage} 

\lettrineabstract{Potential malicious misuse of civilian artificial intelligence (AI) poses serious threats to security on a national and international level. Besides defining autonomous systems from a technological viewpoint and explaining how AI development is characterized, we show how already existing and openly available AI technology could be misused. To underline this, we developed three exemplary use cases of potentially misused AI that threaten political, digital and physical security. The use cases can be built from existing AI technologies and components from academia, the private sector and the developer-community. This shows how freely available AI can be combined into autonomous weapon systems. Based on the use cases, we deduce points of control and further measures to prevent the potential threat through misused AI. Further, we promote the consideration of malicious misuse of civilian AI systems in the discussion on autonomous weapon systems (AWS).}


\section{Why civilian AI should not be neglected}

Autonomous functions and systems disrupt businesses and increasingly enter our daily life. AI from the combination of data-processing and decision-making algorithms with relevant data often serves as the basis for autonomous systems. Usually, the algorithms are not restricted to a particular use case but are flexible in their applications. Because of this characteristic, scholars and AI researchers are more and more concerned about possible usage of AI in criminal or military applications. In August 2017, 116 AI researchers and companies raised their concern on repurposed AI and robotics  and  called in an open letter to the UN Conference of the Convention on Certain Conventional Weapons (CCW) \citep{Letter2017} to ban lethal autonomous weapon systems (LAWS) and a current pledge from the Future of Life Institute against LAWS is open for signature \citep{FutureofLifeInstitute2018}. Further, reports highlight challenges from a militarization of AI \citep{Rickly2017} and the difficulty in restricting proliferation of AWS \citep{UNIDIR2017}.

The current discussion at the Group of Governmental Experts (GGE) about Lethal Autonomous Weapon Systems (LAWS) focuses particularly on autonomous systems that are purposely developed for military applications. However, if expanding the scope to AWS and attacks other than physical, a less clear border between civilian and military AI prevails. Instead, an increased spill-in of autonomous functions from civilian technology into weapon systems can be expected because main research funding lies in the commercial sector \citep{Cummings2017}. The Stockholm International Peace Research Institute highlighted in a report in 2017 that the GGE should investigate "the options for preventing the risk of weaponization of civilian technologies by non-state actors" \citep{Sipri2017}.

With this paper, we aim to show that civilian AI should be included in the discussion on AWS because its misuse by a variety of actors poses severe risks for security. After outlining our definition of an AI system and describing how current AI development works, we distinguish between different modes of AI use. In Section \ref{threats} we show various threats of malicious use of AI and present our three use cases of potential misuse of available AI. Section~\ref{StateEngage} gives implications of misused AI in AWS for states and Section~\ref{prevent} shows possible measures to prevent malicious misuse. Section~\ref{concl} concludes our paper.

\subsection{AI Systems – more than Algorithms}
Even though it might be obvious to assume that autonomous systems based on AI are primarily defined by their algorithms, each system is a complex composition of input data, a predefined goal, the underlying code, and the hardware or software used to interact with either the physical or digital world. A visualization of this can be found in Appendix B.

\subsubsection*{Relevant Data}
The first and most important part for any AI is data. No current AI is able to make meaningful decisions without a certain amount of data that is needed to train and verify the decision-making engine. The required amount of data varies with how well the engine is built and how significant the used data is. The usefulness is determined by noise in the data and the accuracy of labeling. Labels should indicate, if a predetermined goal (e.g. win at a certain game) was achieved and to what degree this was the case (how many points were scored in the game). It is therefore crucial when using already existing data, that it is labeled or when it is collected, that the data is classifiable for an engine in these two categories.

\subsubsection*{Definition of a Goal}
The goals that have been formulated for AI in the past have been rather simple. Even Google DeepMind’s AlphaGo, which in many ways can be considered a complex construct of several decision-making engines, had the clear and simple goal to win at the Game \textit{Go}. However, even complex goals can be broken down into a simple set of tasks and goals, enabling the AI to check in which state it is and which specific goal applies in this case. If there is a large set of variable goals, it is very likely that an own engine would be implemented to define the most important goal in a specific state.

\subsubsection*{Interface}
An interface serves two purposes. The first is to collect data of the current environment and to transmit this data to one or more decision-making engines as input parameters. Without this step, no AI would be able to process the environment it is currently acting in. The second purpose is to perform the action that was decided upon by the software, enabling the AI to engage with its environment as an actor. This interface can either be hardware when interacting with physical objects or additional software when interacting in a virtual environment.

\subsubsection*{Data-Processing and Decision-Making Engine}
The actual core of an AI is its data processing and decision-making engine. Here the AI takes its input parameters, consisting of the values determined by its sensors and usually a processed form of its previous inputs, and makes decisions based on these. For this engine to make useful decisions towards the predetermined goal, it needs access to large amounts of data with which it can be trained and the results verified. Usually, several engines feed each other with input parameters in the execution of complex tasks. Single engines only evaluate a task specific part of the decision. A popular approach to this sort of engine is currently reinforcement learning, although approaches differ according to tasks and goals. Reinforcement learning is a form of machine learning where the acting AI interacts with its environment and tries to maximize the reward that is determined by a function that, if determined correctly, should incorporate all goals. \citep{reinforce}

\subsection{Openness as Key of AI Development}
\label{openness}

Research and development of AI systems are characterized by a high degree of openness, meaning that a resource is mostly free to use, manipulate, and redistribute. This applies to different categories of resources used in the development of AI systems such as hardware and software components, scientific findings and their publications, as well as data required for training of systems. As a result, new achievements in the area often proliferate and diffuse rapidly while further allowing the increase in velocity of innovation.

Openness is not a characteristic of AI development alone, rather it is a result of the openness of the underlying components. Although "open" movements (e.g. open source, open data, open hardware, open access) have been present for many years in various areas, the amount of available resources as well as their accessibility and usability have heavily increased in the last decade. Strong drivers for this development are certainly the increasing worldwide availability of high-speed internet access and growing computing power at sinking costs.
Despite open source software and algorithms being the most commonly known examples for openness, open data and open hardware are equally important drivers for AI development.

\subsubsection*{Levels of Openness}
\begin{figure}[H]
    \includegraphics[width=\linewidth]{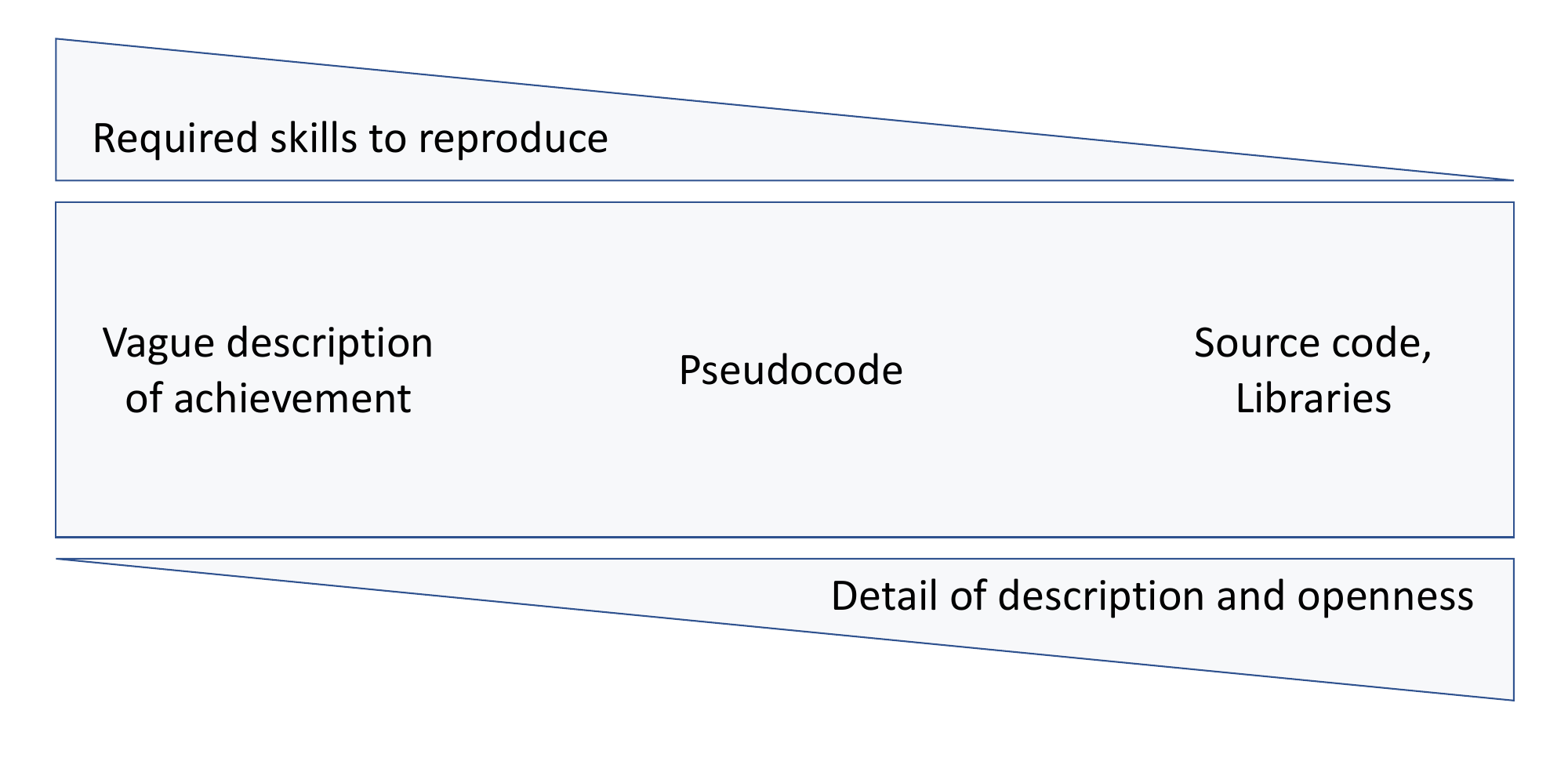}
	\caption{A schematic illustration of the relationship between openness and required skills to reproduce a part of an autonomous system, here depicted exemplary for an algorithm with the skill required to reproduce a functionality (from \cite{Malai2018})}
	\label{fig:AIopen}   
\end{figure}

Openness hast to be viewed on a continuous scale: the level of openness can vary anywhere from a vague or abstract description to fully functional source code, trained models, detailed tutorials, files for 3D printing or full datasets. The usefulness of resources varies together with the openness level: The higher the level of openness for a given resource is, the lower the required skill level and barrier to reproduce or reuse the resource (c.f. Figure~\ref{fig:AIopen}). Generally speaking, the highest degree of openness can be found for software and data while open hardware is merely an emerging phenomenon. However, with the ongoing rise of 3D printing, the impact of open hardware is strongly increased. Just to mention one fascinating glimpse of what could be possible in the future development of open hardware, projects on 3D printed humanoid robots exist. Building instructions and printing templates for miniature robots are already available \citep{3drobot} and at least one project strives for an "Inexpensive 3D Printed Full Size Humanoid Robot" already giving detailed build instructions \citep{fullsizerobot}.

\subsubsection*{Platforms and Sources for Open Content}
Across all openness levels, large online platforms have been key for accessibility and usability of open resources since they provide not only the content (software, hardware, data) itself but allow project management, collaboration across projects and social interactions. Simple user interfaces allow low access barriers and increase the usefulness of available resources. Examples for such platforms are: 
\begin{flushleft}
\begin{itemize}
    \item SourceForge (\url{sourceforge.net}) and GitHub (\url{github.com}) for software
    \item Hackaday.io (\url{hackaday.io}) for hardware
    \item DBpedia (\url{wiki.dbpedia.org}) or Wikidata (\url{www.wikidata.org}) for data
    \item arXiv (\url{arxiv.org}) for scientific publications
\end{itemize}
\end{flushleft}
Furthermore, platforms exist that combine access to multiple resource categories. Kaggle (\url{kaggle.com}), just to mention one, provides a combined development environment and sharing platform for code and data. Additionally, there exist websites simply listing different heterogeneous sources for content (eg. \cite{OpenData}).

Openly available resources on these platforms originate from different sources, namely academia, the private sector, the public sector and the developer-community. Examples include the publication of algorithms on GitHub by university researchers \citep{TUDarmstadt}, open release of software libraries from private companies (e.g. TensorFlow by Google), public data made available by governments, and build instructions on platforms such as Hackaday.io by the developer-community.

\subsection{Malicious Misuse of civilian AI}
Before discussing misuse of AI, it is important to clearly define this topic and differentiate between different modes of AI use. Therefore, we provide in the following a set of definitions and a corresponding categorization scheme for the usage of AI technology.

\subsubsection*{Distinction between Modes of AI Use}
\label{modesAIuse}
In general, technologies can be categorized as military or civilian with respect to the purpose for which they are used. Some technologies can be seen as part of both categories and are then often referred to as dual-use technology. Based on the definitions of dual-use by the Oxford Dictionary\footnote{"[dual use] (of technology or equipment) designed or suitable for both civilian and military purposes."  \url{https://en.oxforddictionaries.com/definition/dual-use}} and the German Federal Office for Economic Affairs and Export Control\footnote{„Dual-use items are goods, software and technology that may be used for civil and military purposes.“  \url{http://www.bafa.de/EN/Foreign\_Trade/Export\_Control/}  \

\hspace{-0.35cm} \url{export\_control\_node.html}}
we come up with the following definition for dual-use AI technologies:\\

\begin{quote}
  \textit{Dual-use AI technologies:} Dual-use AI technologies are those technologies that can have military and civilian use.
\end{quote}

Major funding of development of autonomous systems lies in the commercial sector \citep{Cummings2017} and for a big number of AI algorithms a military use can be imagined. Therefore, the above definition of dual-use of AI technologies seems not very helpful when dealing with potential usage of AI primarily developed for civilian purposes in AWS.

Apart from the \textit{military versus civilian} classification of of AI one can also classify by whether the AI technology is used the way intended by the producer ("intended use") or misused for a purpose unintended by the producer ("misuse"). We therefore define\\

\begin{quote}
\textit{ Misuse of AI:} Misuse of AI is the use of AI for applications that were not intended originally.\footnote{This definition is based upon the definition of misuse by the Cambridge Dictionary: "An occasion when something is used in an unsuitable way or in a way that was not intended". \url{https://dictionary.cambridge.org/dictionary/english/misuse}} \\
\end{quote}

This misuse can either be benign or malicious. We define\\

\begin{quote}
\textit {Malicious use of AI:} Malicious use of AI is the usage of AI technology to an end that threatens security.\\
\end{quote}

\begin{figure}[H]
\begin{centering}
    \includegraphics[width=0.5\linewidth]{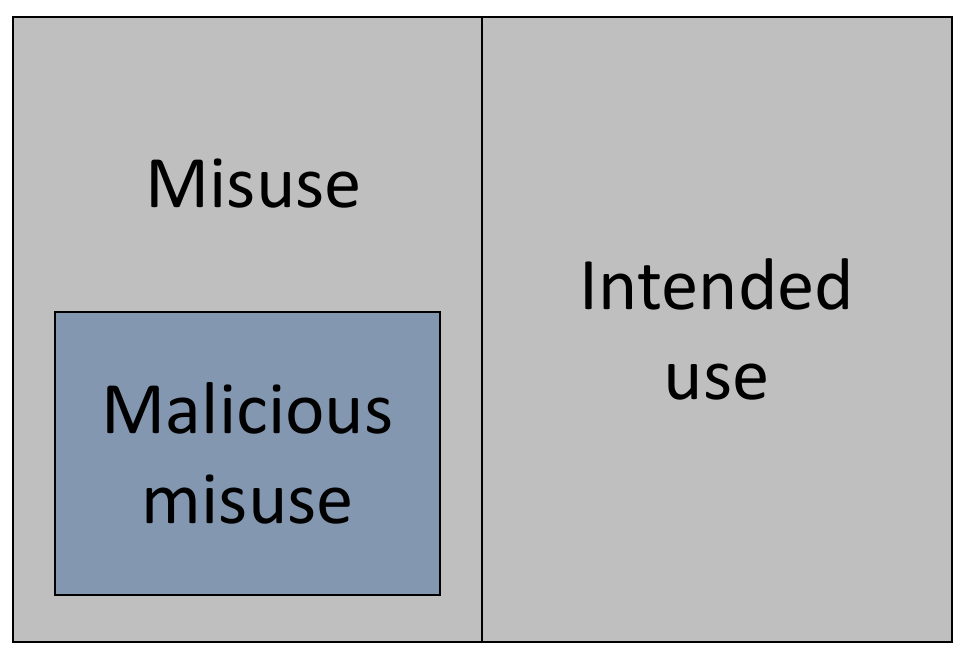}
	\caption{Schematic of AI modes of use with malicious misuse highlighted as being the scope of this paper}
	\label{AIusage}
\end{centering}
\end{figure}

Many innovations are based on unconventional but benign misuse of available and open technology. Therefore, there is misuse of AI components which is benign and an important driver for innovation in the global developer-community. In our work, we do not consider this benign misuse but explicitly focus on malicious misuse of civilian AI, because it threatens security and lacks attention in the current debate. Figure \ref{AIusage} visualizes the different modes of AI use, which define the scope of this paper.

\section{Possible Threats from available AI}
\label{threats}

\subsection{Categorization of Threats}
A prevailing and common association with AWS is that of armed robots and autonomous drones. Aside from these depictions (which most likely arise from successful Hollywood productions), autonomous weapons in virtual environments are similarly threatening security and probably more available than their hardware equivalents. The threats that malicious AI poses, can be assigned to each of the following three categories according to \textit {The Malicious Use of Artificial Intelligence: Forecasting, Prevention, and Mitigation} \citep{Malai2018}.

\subsubsection*{Digital Security}
The first category is digital security threats. The emergence of AI eliminates the currently existing trade-off between scale and efficacy of digital attacks. Potent attacks will no longer be as labor intensive as they are currently by manual work, as an AI will autonomously be performing these complex tasks. Further, autonomous systems themselves (and therefore also AWS) can be be targets. It seems reasonable to expect that digital attacks will be targeting vulnerabilities of AI systems such as data poisoning\footnote{Since most AI systems undergo a training phase that is performed with a set of training data, specifically crafted attack points can be placed in this data. These attack points then possibly transfer into AI systems that were created with the respective data and can then be exploited \citep{poisoning}}.

\subsubsection*{Political Security}
Malicious AI can be used in various ways for attacks when it comes to political security. Possible applications are the automation of surveillance, persuasion through propaganda and deception through manipulation of media. We expect new applications to make use of the classification capabilities of machine learning and analyze human behaviors, mood and beliefs based on available data.

\subsubsection*{Physical Security}
Physical threats are the most prominent of the three, consisting partially of what one might define as classical AWS or LAWS. This can be direct attacks by AI-managed agents that otherwise could not be controlled by humans (such as a swarm of strategically acting drones). However, also attacks that subvert cyber-physical systems and result in threat of physical security (for example an attack on critical infrastructure) should be considered as physical threats of maliciously misused AI.

\subsection{Cases of malicious Misuse}
\label{cases}

Aligned with above-mentioned categories of security threats and the open characteristics of AI development, we developed three exemplary use cases where AI is maliciously misused. We do not consider what kind of attacker uses these AI systems as a weapon, but focus on the feasibility of misuse of available technology. In the remainder of this section we give a high-level description of the three use cases. We attached technical models and an estimation of effort to create the use cases in the Appendix.

\subsubsection*{Digital Security Threat: Social Network Spear-Phishing (Use Case I)}
This use case introduces AI for social engineering cyber-attacks and is based upon work by \cite{seymour_tully}.

It features automated target-user selection and shows the capabilities of neural networks to create natural-language posts that catch the attention of the target-user.
The goal of the AI is to make high-value users of a social network click on an infected link in order to deploy malware and gain access to the system of the target. 
The only hardware required for this use case is a computer system that runs the AI algorithms and that has internet access to the website of the targeted social network.

The first step in generating targeted posts is to collect data about a large amount of users of the network, i.e. their posts and profile information. This is done via scraping scripts. There is a variety of open-source frameworks available for social media scraping\footnote{\url{https://github.com/search?q=social+media+scraping}} or commercial solutions exist that even allow to circumvent ip-address bans such as Crawlera\footnote{\url{https://scrapinghub.com/crawlera}}.

After having collected user data, a clustering algorithm is applied to select a high-value target. For each high-value user that was found, the data collected is analyzed to determine the posting behavior of the user. Both timing of posts during the day and the most frequent topics are part of the analysis.

The topics are used to seed a recurrent neural network that generates target-specific messages and append a shortened malware-link. Although the current state of Natural Language Processing via AI does not allow to always generate fully correct sentences, these messages are much more likely to attract the attention of the target than generic spam, leading to higher click-rates on the infected link. This is partially also promoted by users tolerating a lower standard of grammar and orthography on social networks.

Instead of solely relying on generating target-specific messages via a neural network, one could further increase message quality by incorporating existing posts by others on the topic of interest.

In summary, the usage of AI for social network spear-phishing automates and increases the efficiency of an attack, which used to be considered manual beforehand.

\subsubsection*{Political Security Threat: Propaganda through Deepfakes (Use Case II)}
This use case introduces a scenario for manipulation of public opinion based on modern AI technology. It leverages 3D face reconstruction methods and neural networks to impose desired speech as well as facial and body expressions onto the portrait video of a desired target person. The resulting output video, commonly referred to as "Deepfake", will be almost indistinguishable from genuine sources without the need for manipulation by hand and can also be generated in real-time. This enables new types of attacks that would not be realizable without AI technologies.

At the core of this use case is a software that takes the portrait video of a source actor as input and transfers (in real-time) the head- and torso-movements, facial expressions and speech onto a target person, generating a photo-realistic manipulated output video and convincing impersonated audio. Different approaches exist to implement this use case and the detail and amount of controllable features of the target in the output video varies. The following description is based on the approach in \citep{Kim2018} for the video and \citep{VoiceGAN2018} for impersonation of the targets voice.

The face-manipulation software consists of three main parts: A 3D parametric face reconstruction of both source and target from the videos, the reconstruction of a synthetic target video where head parameters are transferred from the source and finally the prediction of a photo-realistic output video from the synthetic target video using a generative neural network. The neural network also takes care of estimating torso and background shadows given the desired head-pose for the target. The voice impersonation relies on a generative neural network that, from an input audio source, generates a audio sequence mimicking the voice style of the target person.

The training of the utilized neural networks requires short portrait video sequences of both source actor and target person with appropriate lighting and a static background as well as corresponding audio sequences comprising speech samples. Given the wide availability of content on online video platforms the required video and audio for a large number of desirable target persons can be assumed to be easily retrievable. 

Given the published results of these two approaches combined in this use-cases, various scenarios are possible where the contained technologies are maliciously misused. The target person will usually be a person of public interest with a large influence on public opinion. Distribution channels could be online video platforms providing user-content (e.g. YouTube) or editorial content by media companies as well as TV broadcast. The goal of such attacks can range from the dissemination of propaganda to the manipulation of elections or societal destabilization.

\subsubsection*{Physical Security Threat: Strategically acting Swarm (Use Case III)}

This use case exploits civilian strategic AI for military planning and tactics. It shows the capabilities of reinforcement learning from a gaming AI to control swarms of autonomous robots on the battlefield. The goal of this type of AI is to fight on an unknown battlefield optimally with respect to minimal own losses but the highest number of assassinated enemies possible. 

The autonomous robots of this use case are equipped with optical sensors (cameras, ultra sonic or LIDAR), propulsion technology for movement (on the ground as for example humanoid robots or flying drones) and an execution agent with the capacity to attack a target. The execution agents can be small weapons mounted to the movement hardware, explosive devices or the agent itself when attacking by direct contact. Hardware for autonomous robots (eg. \url{https://www.robotshop.com/} and explanations for building autonomous drones (eg. \cite{Burkle2011}) are abundantly available.

The algorithm to control the movement of the swarm of agents, is threefold. First, the orientation movement of the agents in the unknown battlefield needs to be ensured. This can be achieved by dynamically updating the maps while the agents are moving and observing their environment with their sensors. Simultaneous Localization and Mapping (SLAM) is typically applied by self-driving vehicles and drones and fully-functional libraries and source code is publicly available \citep{SLAM2011, TUDarmstadt}. SLAM works in a decentralized manner, therefore no central agent or communication possibilities to the battlefield need to exist. 

To further ensure no collision of the drones with each other, sophisticated approaches for swarm formation of robots and drones exist either with central \citep{Alonso2015} or distributed control by self-organization of the autonomous agents \citep{AlonsoMora2016} which show impressive performance even by navigating through narrow passages. 

At last, a general trajectory for the movement of the whole swarm and attacking the enemy needs to be created by an algorithm. Because no knowledge on where the enemies are located and what actions they undertake exists, this problem can be categorized as an incomplete information game. Recent advances in reinforcement learning show promising results in winning against humans in these games \citep{Neller2013, Hartley2017}. To take one example, the gaming AI Libratus defeated world-class poker players by out-bluffing them \citep{Spice2017}. In order to achieve this, Libratus predicted the actions of the other human players. It uses game moves as basic blocks, predicts enemy behavior and evaluates the outcome of different strategies consisting of these basic blocks under probable enemy behavior.

This AI could be adopted to military strategy creation by using military tactics instead of game moves and their real-time evaluation. Tactical basic blocks such as distraction, deception and counter-offensives would be evaluated based on their possible impact under consideration of possible actions of the enemy. Since the autonomous agents could be designed to be able to communicate in real-time with each other, the human enemies would have severe disadvantage against the highly agile agents.

\section{Why States should engage}
\label{StateEngage}

As seen in the previous section, AI systems have the potential to bring dramatic, systematic changes in security.
In this section, we present arguments from \cite{Malai2018} and relate them with implications deduced from our use cases.

\subsection*{Expansion of existing Threats}
If AI technology is diffused, its properties of efficiency and scalability will make existing attacks possible
\begin{itemize}
    \item for more actors to carry out,
    \item on a wider scale,
    \item on more targets.
\end{itemize}
This leads to a power shift from governments and big organizations to non-state actors, as large resources are no longer necessary for an attack. The above presented cases of AI application threatening digital and physical security by spear-phishing (use case I) and the autonomous-agent swarm (use case III) illustrate this novelty. In many occurrences where hackers have gained access to critical infrastructure or persons, manual spear-phishing was used, which requires a large amount of resources. Aerial drone attacks used to be carried out by large military powers only, as drone development was very expensive. In both cases, the implementation of AI can reduce costs drastically and make it available to more actors, e.g. non-state actors.

\subsection*{New Attacks}
AI can bring about new attacks, that formerly were not possible. An example is presented with the deepfake AI application (use case II). The capabilities of AI exceed humans by far in certain areas, such as mimicking others. This leads to serious threats of political security by misinformation and impersonation. 

Apart from the diffusion of AI itself, the diffusion of interfaces, particularly hardware through 3D printing, also poses threats to physical security. Combining maliciously misused algorithms with easily adjustable hardware components via computer aided modeling software allows to build malicious autonomous systems which are able to execute physical attacks without human intervention.

\subsection*{Changed Character of Attacks}
AI-enabled attacks will not only be more effective, but also allow for a finer targeting and more difficult attribution of the attack. An example is shown above with the autonomous swarm AI application (use case III), which can be deployed to assassinate a specific person while at the same time making it unnecessary for the attackers to be present or even executing the action. This further results in a psychological distance and might increase the likeliness to engage in a violent conflict.

\subsection*{Interest of States}
It is in the original interest of states to protect digital, political and physical security of their infrastructure, political system and population. As outlined above, this is threatened already by diffusion of civilian AI. Furthermore, a power shift from governments to non-state actors through potentially misuse of diffused AI in a malicious way will fuel asymmetric conflicts worldwide – further decreasing geopolitical stability. For these reasons states should engage as of today and take action to prevent AI-supported attacks.

\section{Prevention of malicious Misuse}
\label{prevent}
\begin{figure}[H]
    \includegraphics[width=\linewidth]{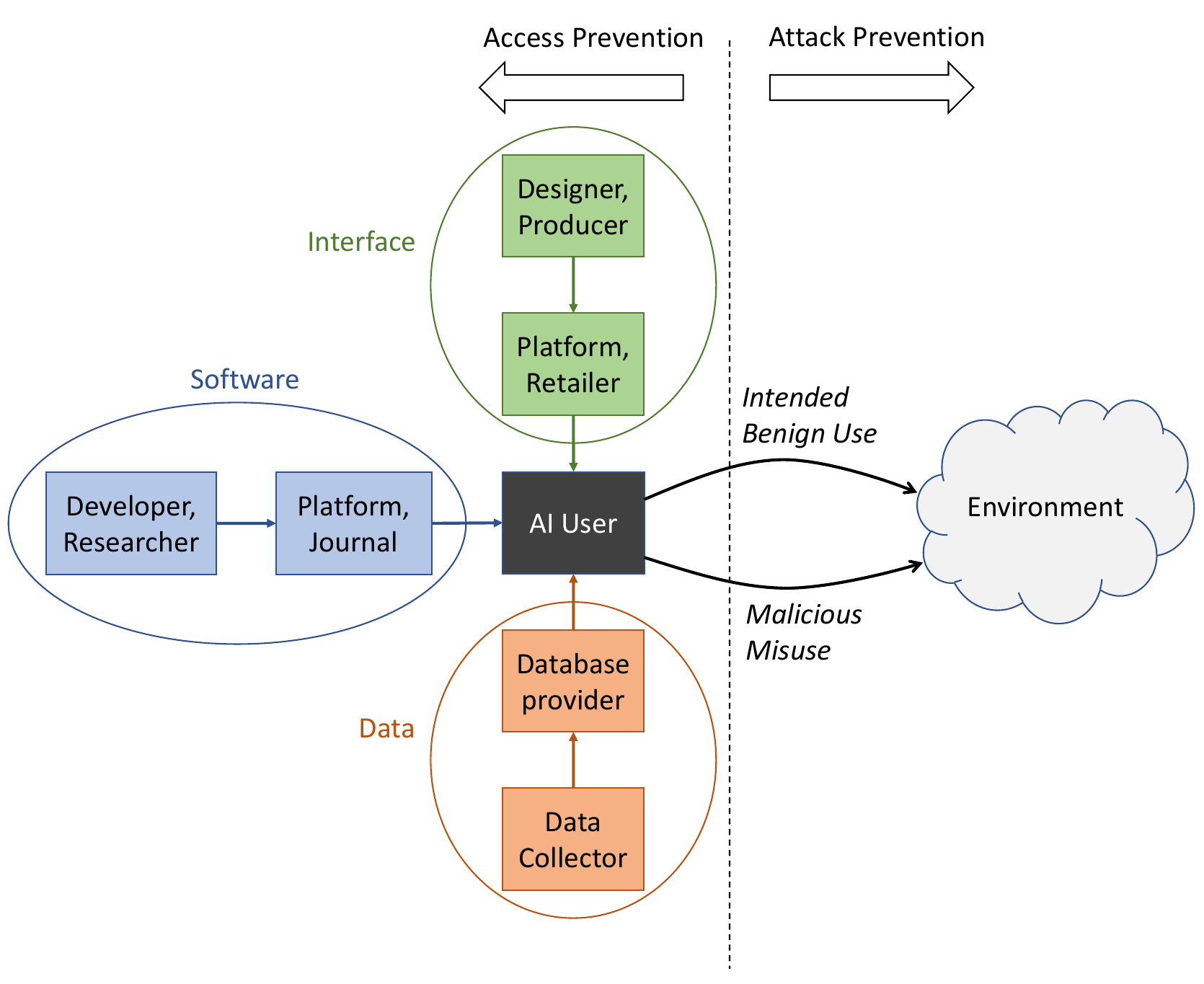}
	\caption{Diffusion and proliferation of AI systems with access and attack prevention as possible measures against malicious misuse}
	\label{aidiffusion}   
\end{figure}

Malicious misuse of AI can be prevented by restricting access to and diffusion of AI functions that can be misused in a malicious way. However, it is also important to prevent attacks of potentially malicious systems. Therefore, both stages need to be considered and we give possible measures for preventing diffusion and attack in this section. Figure~\ref{aidiffusion} shows an illustration of the proliferation chain of AI systems from the sources over a potential (mis-)user to the environment in which an autonomous system is used.

\subsection{Access Prevention through Points of Control}
As outlined in Section \ref{openness}, openness of software, data and hardware components is a key characteristic of civilian AI system development and basis for the fast-moving innovation in this field of technology. On the other hand, preventing access of malicious actors to AI systems that can be misused is highly important to ensure digital, political and physical security as shown in Section \ref{StateEngage}. We therefore propose a threefold approach to balance the seemingly contradicting interests. A visualization of these points of control can be found in Appendix A. 

\begin{itemize}
    \item AI system components could be classified as \textit{critical/non-critical} with respect to malicious misuse before they are made available on online platforms.
    \item Users that want to download and reuse critical components need to register at the platform and follow a certification process in which they are given a \textit{trusted/non-trusted} status. If a user accesses the components, a watermark is put into it in order to to track non-authorized proliferation. 
    \item AI system components are available completely and in full resolution for trusted users and partially or in a reduced resolution (sub-sampling) to non-trusted users.
\end{itemize}

	\subsubsection*{Data}
	Data is a crucial part to developing every AI, so limiting access to sensitive data seems like a logical consequence. Depending on the mathematical functions that are performed on the data either \textit{somewhat homomorphic encryption} or \textit{fully homomorphic encryption} could be applied, allowing the encryption of data before or after it is analyzed or processed. This would enable the scientific or open-source community to still demonstrate the success of their new machine learning algorithms, without having to fully disclose what data they were using and only publishing unproblematic or aggregated data in a decrypted state \citep{fhe}. Research in this area is ongoing but we consider this a possible method of ensuring control and privacy of data. 
	
	Another simpler measure would be to reduce the information contained inside datasets  by sub-sampling. Here sensitive information would either be rounded or completely omitted. This could take shape in the form of anonymizing datasets that belong to potential targets or reducing the fidelity of saved GPS coordinates.

	\subsubsection*{Hardware} 
	Historically, diffusion of dual-use goods and hardware was controlled by multilateral proliferation and trade regimes. As outlined in Section \ref{modesAIuse} it seems challenging to clearly define dual-use AI technology because a number of technologies can be used in military context. To our knowledge, most current trade and proliferation regimes only have implicit regulation on components for autonomous systems as novel technology.
	
	As a first step, an approach for the classification of AI technology should be developed and continuously reviewed on potential harm. This could be done by a committee of robotic and AI experts regularly briefing a multilateral committee.

    Another problem poses the emergence of 3D printing. National borders and export control do not effectively prevent that templates with weapons can be produced at home without any knowledge of who created which object. This year, cases of 3D printable guns drew public attention. However, it probably only marks the beginning of a more fundamental change of the challenge to prevent diffusion of harmful technology. Concrete steps should be undertaken to discuss liability issues of platforms where designs and templates for malicious 3D products are shared. Furthermore, the possession of such drafts could be prohibited. As discussed above, signatures for tracing the diffusion could apply also for these templates.

	\subsubsection*{Software}
    In \ref{openness} several common platforms essential to the diffusion of AI software components such as GitHub have been described. One measure to prevent uncontrolled access would be demanding users of such platforms to register if they want access to sensitive information. This could be done with a name and an address, a unique digital identifier or passport data. Gating access to the code this way would allow for several subsequent stages of revealing of personal data, making it possible to choose the corresponding stage according to the degree of possible misuse of the code. This would not be too different from publishers or scientific journals that provide access to their databases through various different licensing models. Registered users would be incentivized to not misuse the code they acquire and unregistered users would no longer have access to sensitive program code. In cases of highly misusable code, we believe that demanding certification for ethical coding or trustworthiness might be reasonable a requirement, although the effectiveness of a measure like this would have to be explored.

\subsection{Measures for Attack Prevention} %
Preventing the access to sensitive information and AI that could potentially be misused, is only one measure to prevent malicious use. It is of equal importance to ensure that countermeasures exist once malicious AI has been created and is being used. 

One way to do this is to employ AI to classify and detect abnormalities (and potentially malicious software, forgeries and social bots), as the IT-security community is already doing. Other approaches from the field of IT-security, such as the responsible disclosure of vulnerabilities and automatic updating, could easily be applied to AI as well \citep{Malai2018}.
Internationally, a body for assistance and exchange of best practices between states could be promoted. This would also help regions with less experience in AI and its development to protect against attacks by AI systems.

Moreover, it could be discussed if certain functions of civilian autonomous systems and hardware should be prohibited. This is already done by included no-fly zones in drones where certain prohibited areas cannot be entered. Finally, it can be imagined that a remotely-controlled functionality for an emergency stop needs to be included into an AI by law.

\subsection{Further Measures}
Besides access and attack prevention, further non-technical measures seem promising to prevent threat from misused AI. The most important one is to start transdisciplinary discussion about malicious misuse of available AI for autonomous weapon systems. To achieve this, collaboration between different UN bodies, technical experts from academia and the private sector, states and the civil society should be promoted. This could be formalized through the establishment of periodically meeting working groups and a permanent multidisciplinary UN-committee on (L)AWS.

In particular, the inclusion of the civil society in the discussion on AWS and implications of AI in general is of high importance. Besides stronger legitimization of the discussion on this issue, informed and sensitized citizens are the most powerful measures to ensure political security even with attacks from malicious actors. Additionally, other regimes such as trade, labor and non-proliferation of weapons should be encouraged and supported when discussing how to deal with AI and therefore actively shape the future of this powerful technology.

As last step, a stronger promotion for forming an international codex for AI developers and sensitizing them about societal impact of their work could also help in restricting malicious misuse of available AI.

\section{Conclusion}
\label{concl}

Resulting from the open and collaborative development of autonomy, software for artificial intelligence and other components of autonomous systems such as training data and advanced robotics are often available and easily accessible online. This allows innovations and development of AI at a high pace. However, we show in three use cases that these accessible systems and particularly AI can be maliciously misused and threaten physical, digital and political security.

In order to limit threat from the malicious misuse of available AI, while at the same time ensuring openness and international AI development in the future, a collaborative approach seems necessary. We encourage states to take actions today to prevent AI-based attacks on large scale, national unilaterism and internet restriction in the future.

Further, we suggest in this work to monitor and prevent access to potentially malicious technology and data through points of control in the diffusion chain of AI technology. Technical measures for attack prevention of maliciously misused AI can be imagined, however need further research. We motivate a transdisciplinary approach to solve the issue by bringing together governments, ethicists, lawyers, AI and robotics experts, both from the private sector and academia, and the civil society.

\subsection*{ConsciousCoders}

\begin{wrapfigure}{l}{0.2\textwidth}
\includegraphics[width=0.9\linewidth]{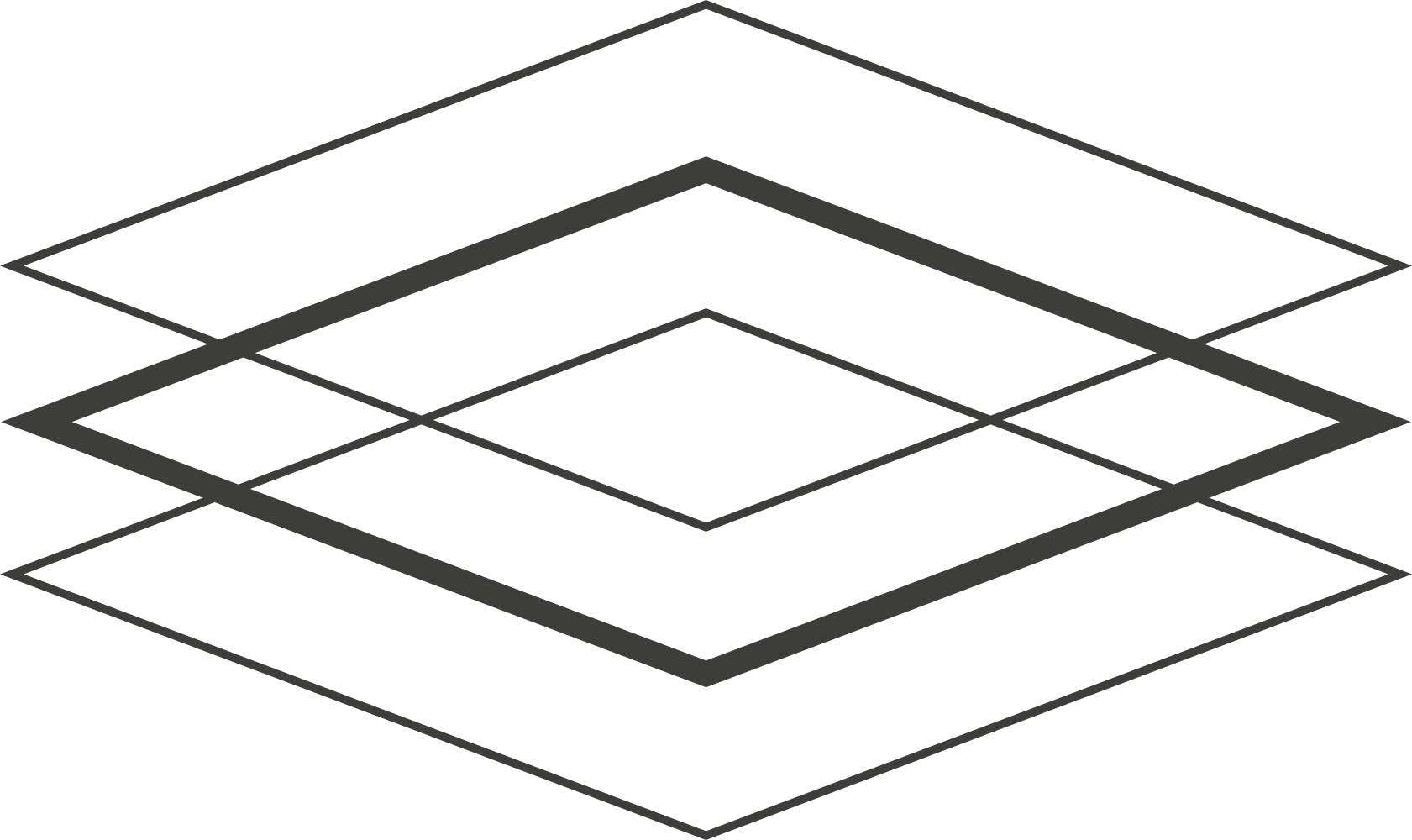}
\caption{Logo of ConsciousCoders}
\label{fig:subim1}
\end{wrapfigure}

 ConsciousCoders is a Munich-based initiative of students and young professionals to discuss the impact and shape the future of artificial intelligence and data science. The project group "Misuse of AI" aims at contributing a technological viewpoint to the discussion around autonomous weapon systems. Moreover, we motivate developers to reflect their work in order to achieve beneficial AI. More information about the initiative can be found at \url{https://www.consciouscoders.io}.


\newpage
\printbibliography[title={Bibliography}] 


\onecolumn
\section*{Appendix}

\subsection*{A) Points of Control for Access Restriction to AI System Components}
\label{App4}

\begin{figure}[H]
    \centering
    \includegraphics[width=.9\linewidth]{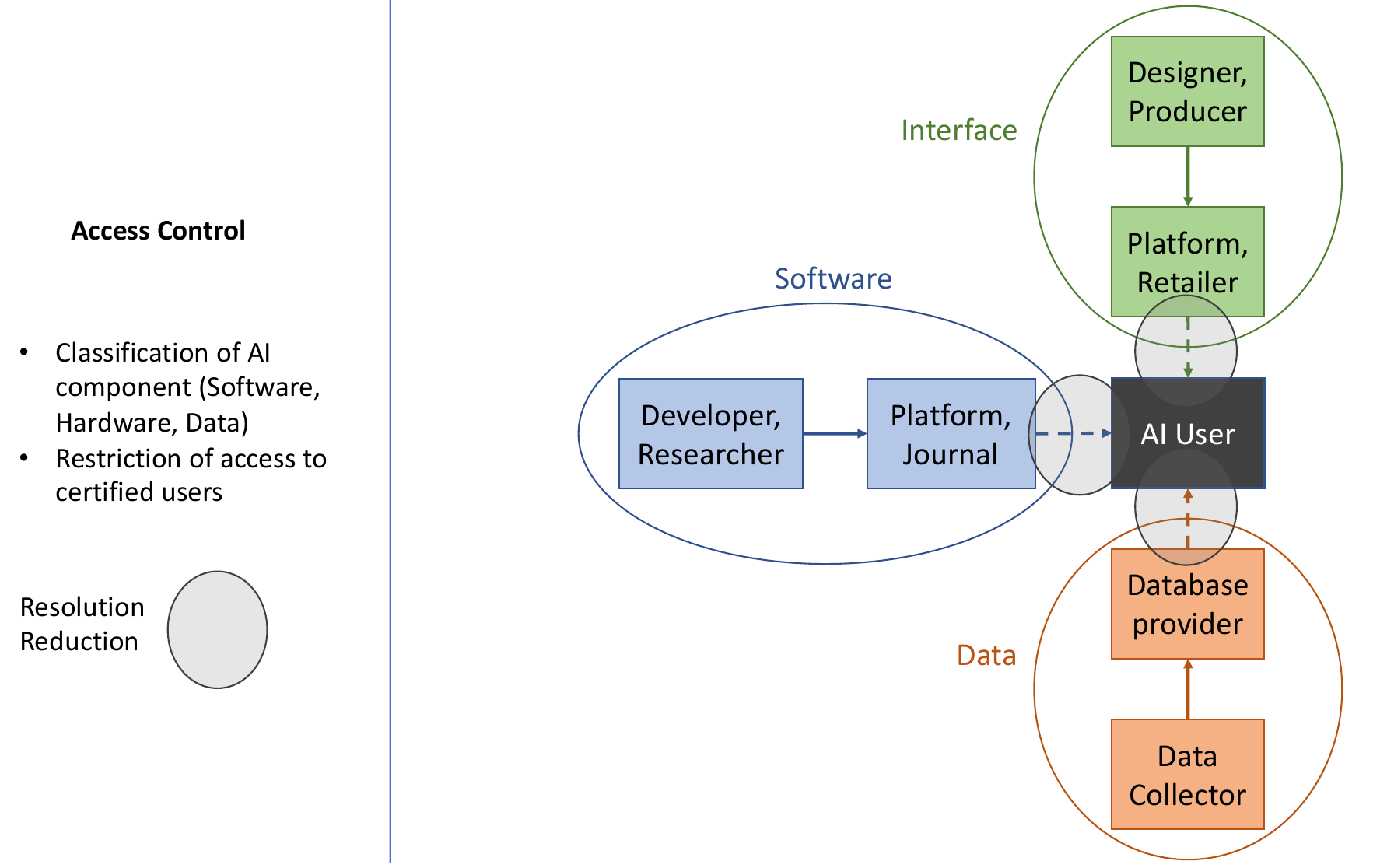}
    \caption[]{Diagram showing access of uncertified users restricted to down-sampled (resolution reduction) AI components}
    \label{fig:restricted_access}
\end{figure}

\subsection*{B) Schematics of Autonomous System}

\begin{figure}[H]
    \centering
    \includegraphics[width=.9\linewidth]{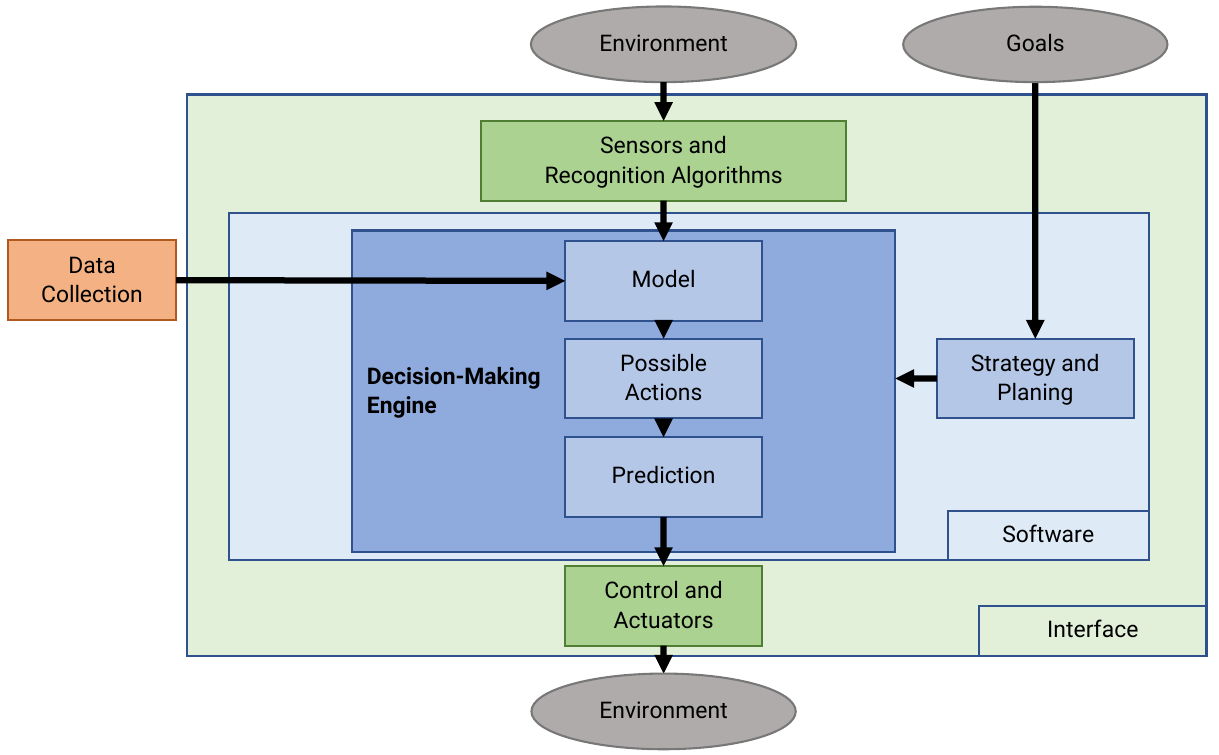}
    \caption[]{Our proposed schematic of an autonomous system based on AI}
    \label{fig:restricted_access2}
\end{figure}

\subsection*{C) Technical Details of Use Cases}

\subsubsection*{Use Case I: Social Network Spear-Phishing}
\label{App1}

Estimated technical feasibility of the use case:
\begin{itemize}
    \item Scraping scripts are open source and therefore available, although the code does need to be adapted.
    \item Clustering algorithms are part of statistic packages for developer environments. A little expertise is required to select the right clustering algorithm for the data at hand.
    \item The implementation of a recurrent neural network requires some expertise, but little actual work, as frameworks and tutorials are available. 
\end{itemize}

Therefore, the skills required are roughly those of an undergraduate level of computer science.

\begin{figure}[H]
    \centering
    \subfigure[Model]{
        \includegraphics[width=0.89\linewidth]{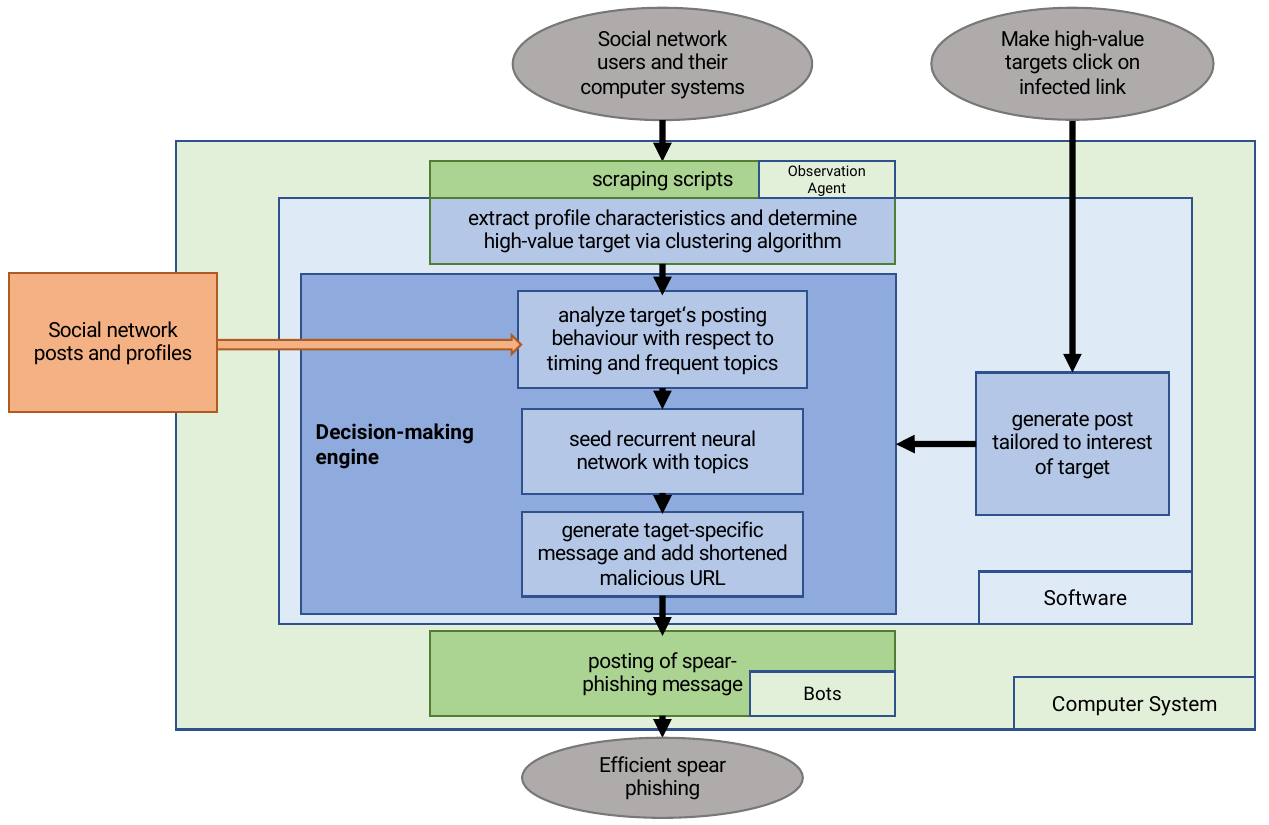}
        \label{fig:TM1}
    }
    \subfigure[Openness estimation]{
        \includegraphics[width=0.6\columnwidth]{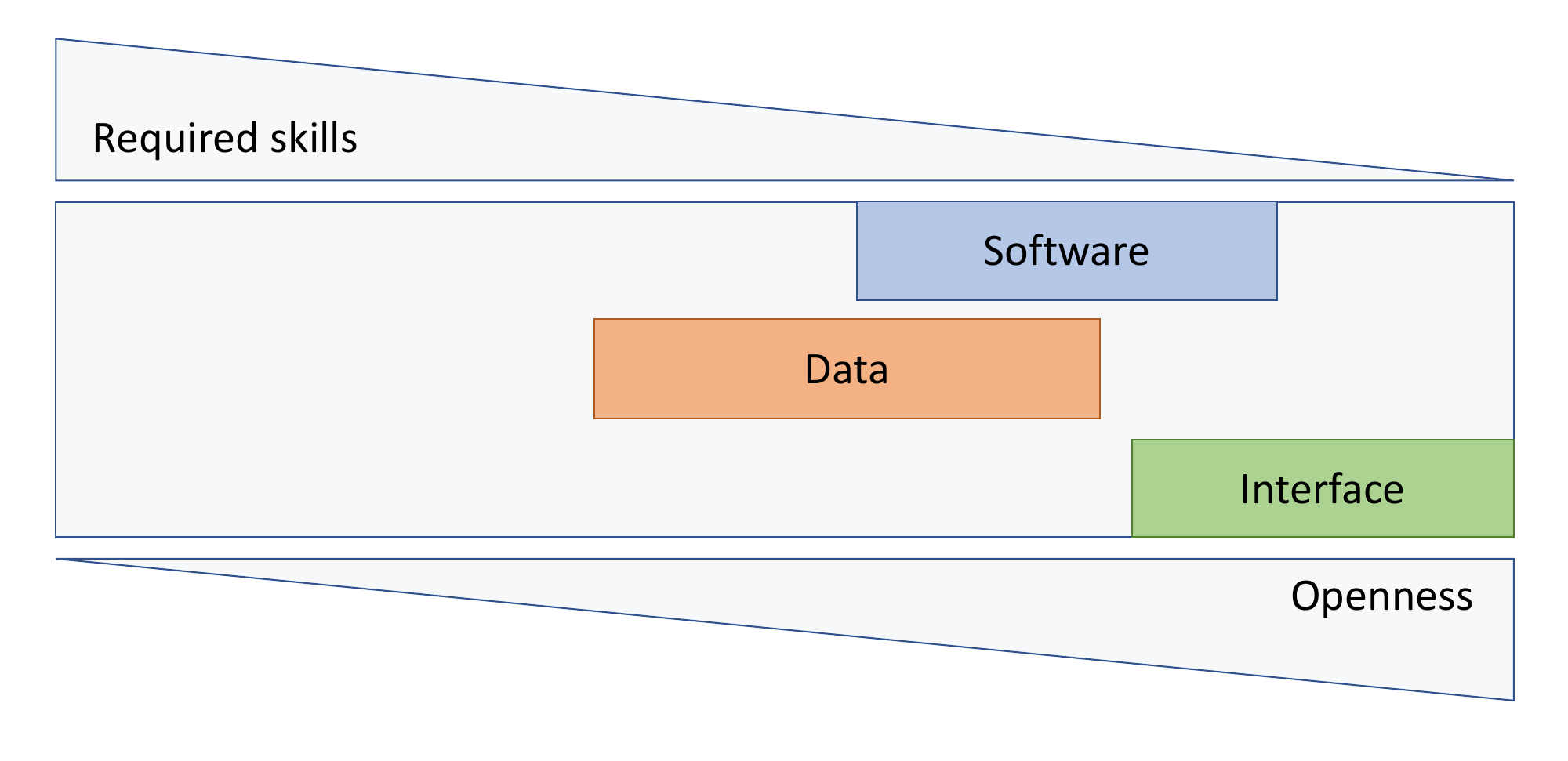}
        \label{fig:Open1}
    }
    \caption[Use Case I]{Detailed technical model of use case I together with an estimation of the openness of corresponding data, software and the interface}
    \label{fig:Technical1}
\end{figure}
	
\newpage	
\subsubsection*{Use Case II: Propaganda through Deepfakes}
\label{App2}

Estimated technical feasibility of the use case:
\begin{itemize}
    \item The necessary videos of the target person can usually be found online or recorded from a TV broadcast because we assume that the person is of public interest.
    \item Sourcecode for the face reconstruction is available in an older stage\footnote{Denoising autoencoder + adversarial losses and attention mechanisms for face swapping \url{https://github.com/shaoanlu/faceswap-GAN}}, for the mentioned accurate version with body reconstruction only a rough description is published.
    \item Voice generation is available even by a number of commercial providers from for example Lyrebird\footnote{\url{https://lyrebird.ai/}}.
    \item The required interface to broadcast the created propaganda videos is either access to online streaming channels or TV broadcasting which differ in their openness.
\end{itemize}
Therefore the level of skills required is at the level of a graduate computer science student.

\begin{figure}[H]
    \centering
    \subfigure[Model]{
        \includegraphics[width=0.89\linewidth]{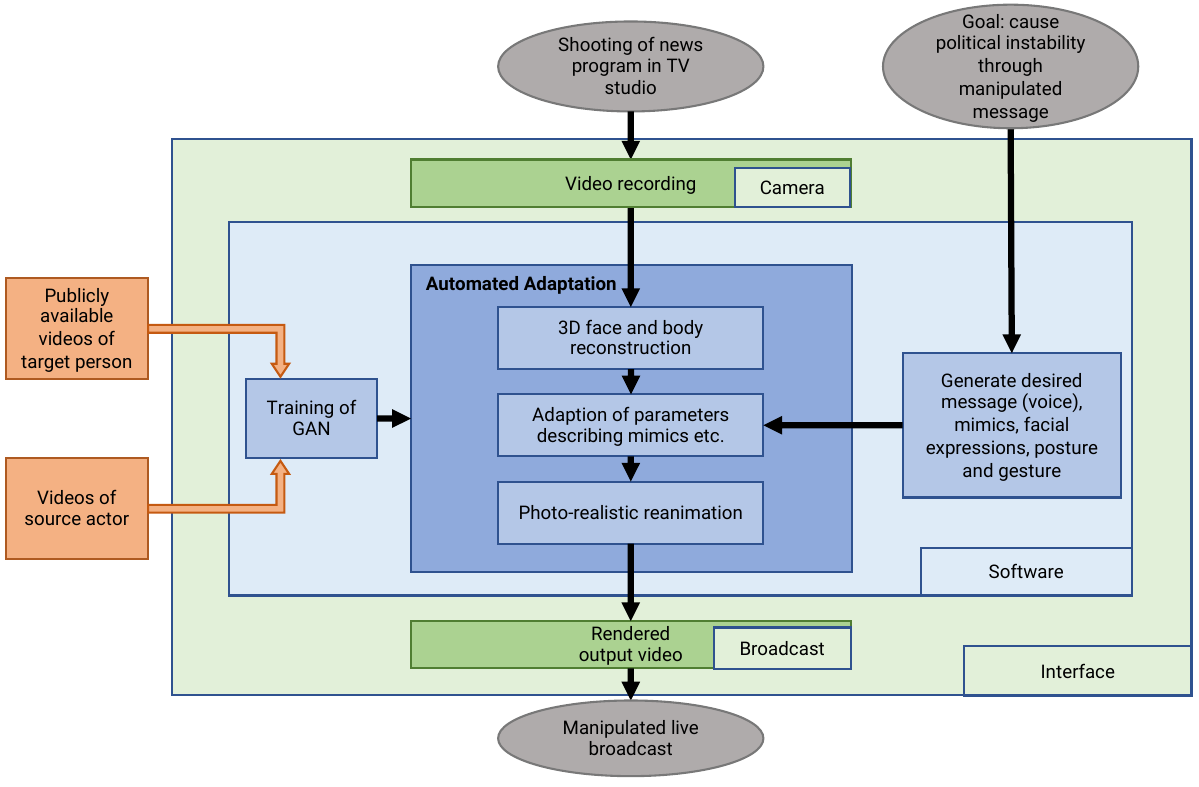}
        \label{fig:TM2}
    }
    \subfigure[Openness estimation]{
        \includegraphics[width=0.6\columnwidth]{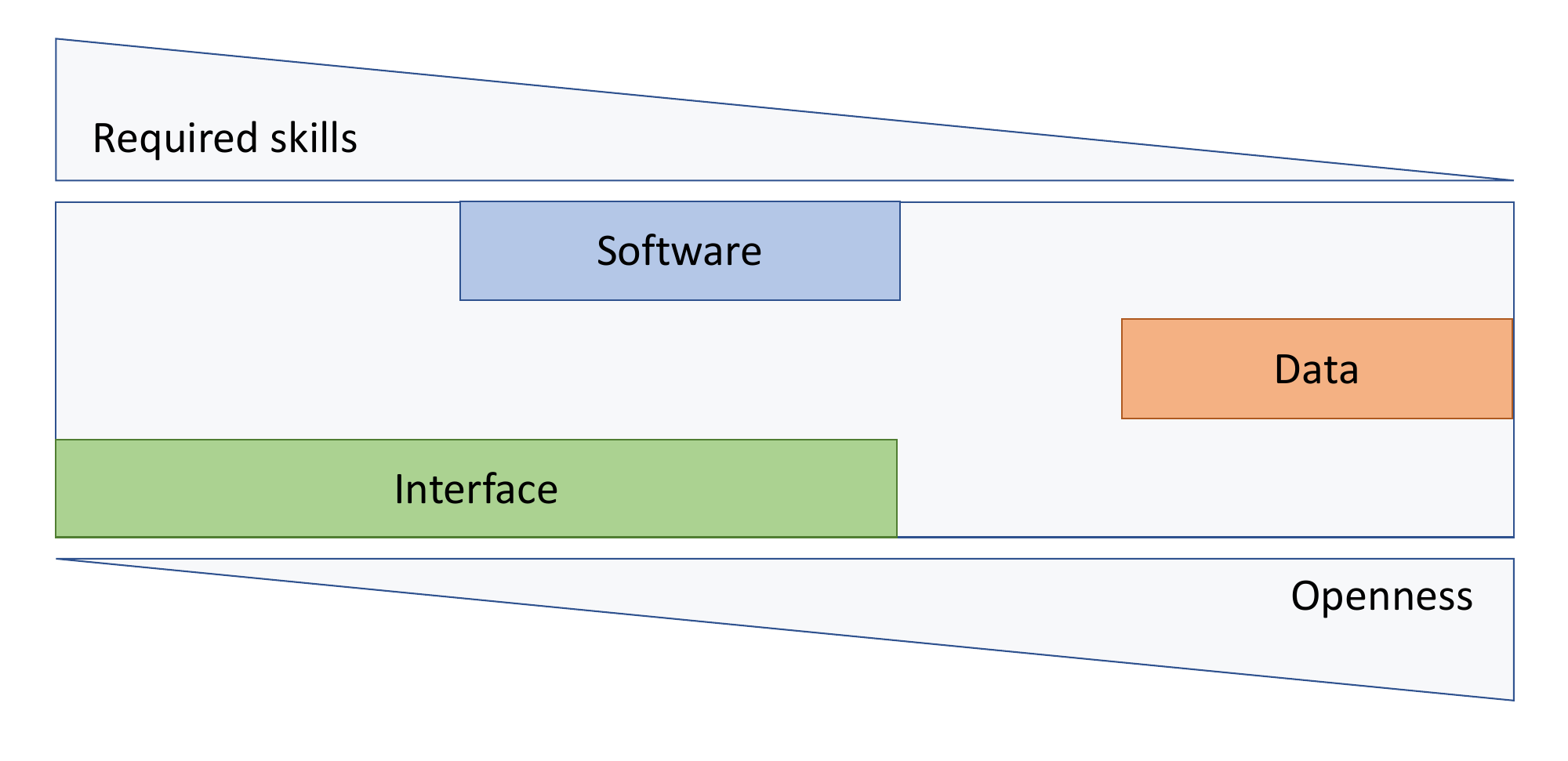}
        \label{fig:Open2}
    }
    \caption[Use Case II]{Detailed technical model of use case II together with an estimation of the openness of corresponding data, software and the interface}
    \label{fig:Technical2}
\end{figure}
	
\newpage	
\subsubsection*{Use Case III: Strategically acting Swarm}
\label{App3}

Estimated technical feasibility of the use case:
\begin{itemize}
    \item Hardware for the moving agents and explanations how to use it in autonomous systems can be found on the internet
    \item Execution agents for attacking the enemy are usually non-civilian and require more effort to get
    \item The algorithms for map creation and collision-free movement of the agents are available in ready-to-use packages
    \item Algorithms for strategy creation can be found as source codes for other incomplete information games such as Poker, to adopt it to a battlefield, a lot of effort is needed. This might change in the future as DeepMind and others are working on AIs that can consistently beat human players in games, where a lot of situations can be applied to real battlefields (such as StarCraft II) \footnote{"DeepMind and Blizzard open StarCraft II as an AI research environment" \url{https://deepmind.com/blog/deepmind-and-blizzard-open-starcraft-ii-ai-research-environment/}}.
\end{itemize}
Therefore, level of skill required: Graduate or PhD level of Electrical Engineer or Computer Scientist.

\begin{figure}[H]
    \centering
    \subfigure[Model]{
        \includegraphics[width=.89\linewidth]{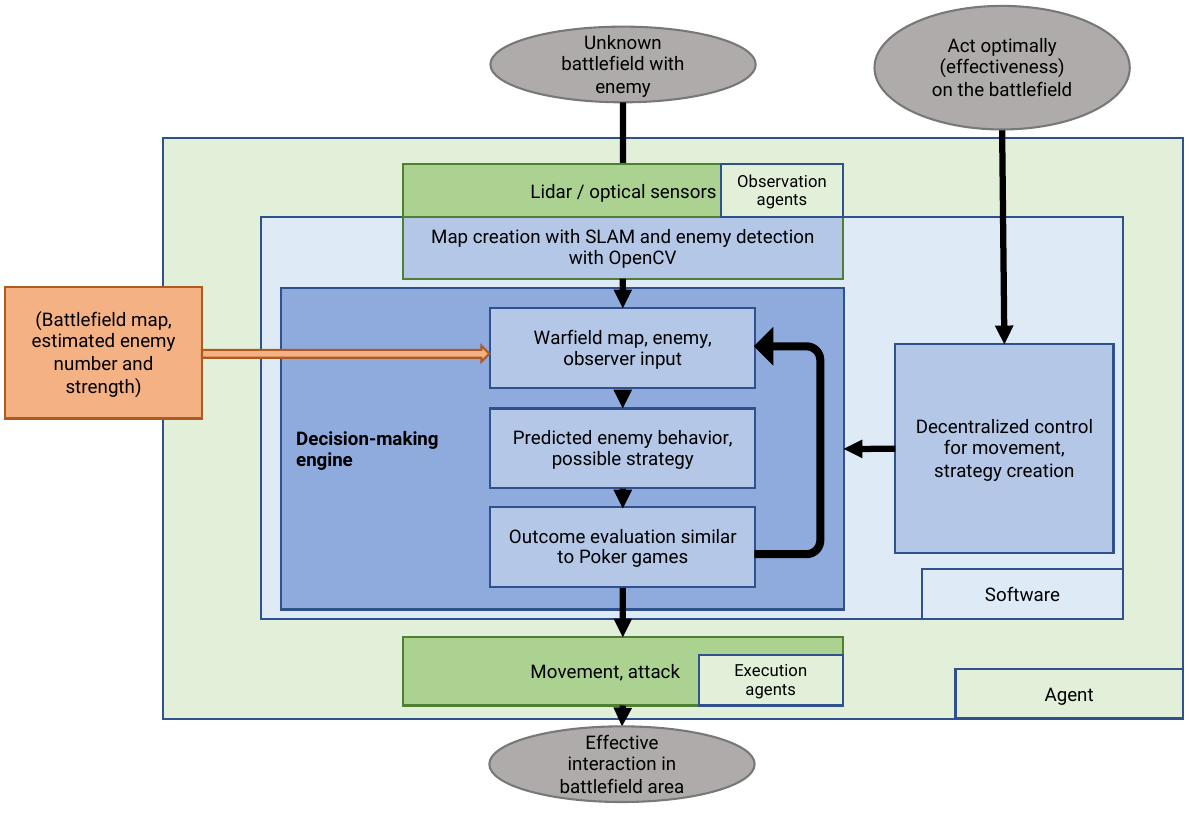}
        \label{fig:TM3}
    }
    \subfigure[Openness estimation]{
        \includegraphics[width=0.6\columnwidth]{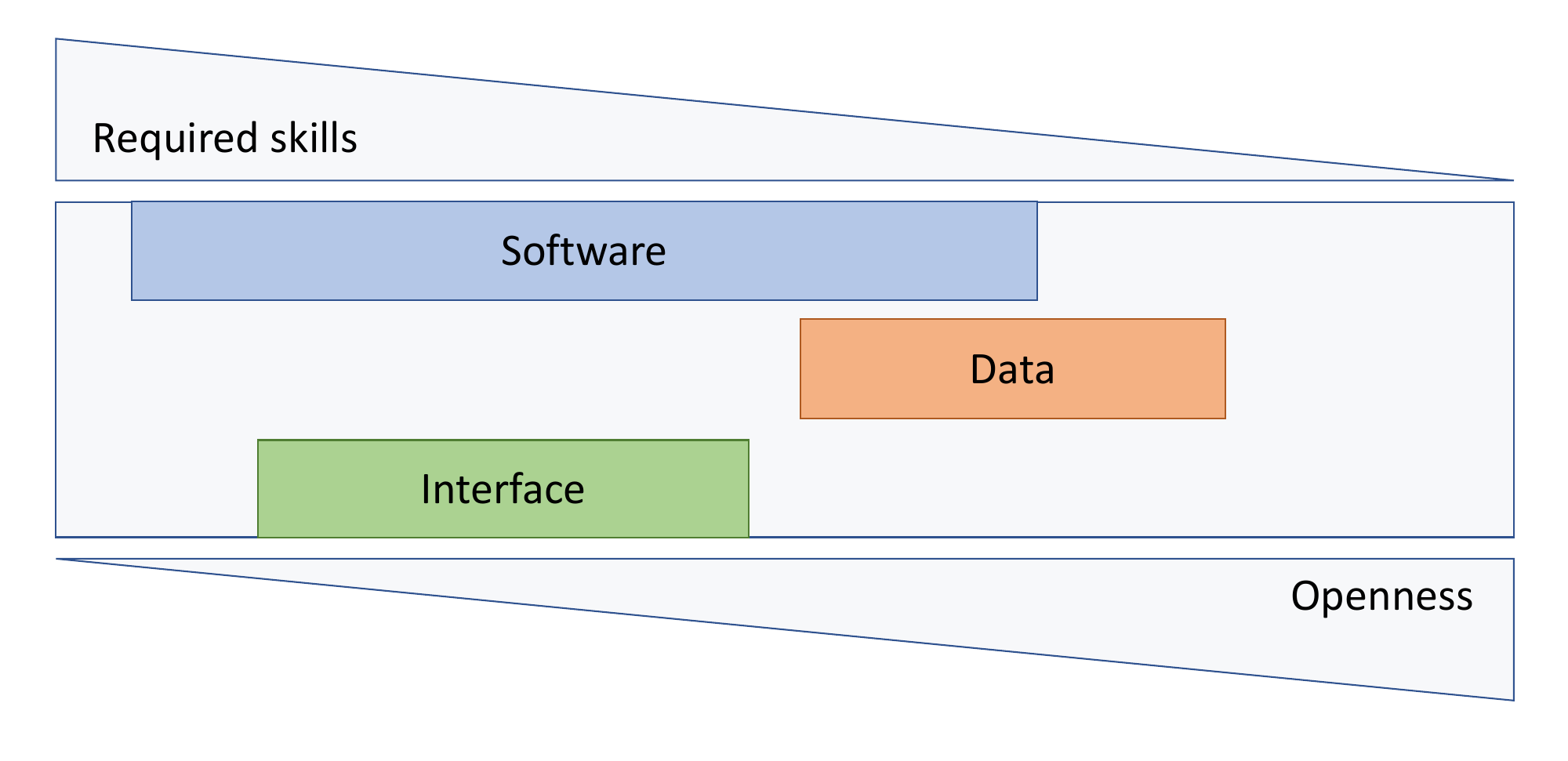}
        \label{fig:Open3}
    }
    \caption[Use Case III]{Detailed technical model of use case III together with an estimation of the openness of corresponding data, software and the interface}
    \label{fig:Technical3}
\end{figure}

\end{document}